\documentclass[twocolumn,showpacs,preprintnumbers,amsmath,amssymb,prb]{revtex4}
\usepackage{graphicx}% Include figure files
\usepackage{dcolumn}% Align table columns on decimal point
\usepackage{bm}% bold math
\usepackage{tabularx}% Table width
\usepackage{amsmath}% Higher math
\begin{document}

\title{Carrier dependent ferromagnetism in chromium doped topological insulator $Cr_{0.2}Bi_xSb_{1.8-x}Te_3$}

\author{Bin Li}
\affiliation{Department of Physics, State Key Laboratory of Surface Physics, and Laboratory of Advanced Materials, Fudan University, Shanghai 200433, People's Republic of China}
\author{Qingyan Fan }
\affiliation{Department of Physics, State Key Laboratory of Surface Physics, and Laboratory of Advanced Materials, Fudan University, Shanghai 200433, People's Republic of China}
\author{Fuhao Ji}
\affiliation{Department of Physics, State Key Laboratory of Surface Physics, and Laboratory of Advanced Materials, Fudan University, Shanghai 200433, People's Republic of China}
\author{Zhen Liu}
\affiliation{Department of Physics, State Key Laboratory of Surface Physics, and Laboratory of Advanced Materials, Fudan University, Shanghai 200433, People's Republic of China}
\author{Hong Pan}
\affiliation{Department of Physics, State Key Laboratory of Surface Physics, and Laboratory of Advanced Materials, Fudan University, Shanghai 200433, People's Republic of China}
\author{S. Qiao}
\email{qiaoshan@fudan.edu.cn}
\affiliation{Department of Physics, State Key Laboratory of Surface Physics, and Laboratory of Advanced Materials, Fudan University, Shanghai 200433, People's Republic of China}
\date{\today}

\begin{abstract}
Carrier-independent ferromagnetism of chromium doped topological insulator  $Bi_xSb_{2-x}Te_3$ thin films,which cannot be explained by current theory of dilute magnetic semiconductor, has been reported recently. To study if it is related to the distinctive surface state of topological insulator, we studied the structural, magnetic and transport characters of $Cr_{0.2}Bi_xSb_{1.8-x}Te_3$ single crystals. The Curie temperature $T_c$, which is determined from magnetization and anomalous Hall effect measurements by Arrott plots, is found to be proportional to $p^{1/3}$, where p is the hole density. This fact supports a scenario of RKKY interaction with mean-field approximation. This carrier density dependent nature enables tuning and controlling of the magnetic properties by applying a gate voltage in the future science researches and spintronics applications.
\end{abstract}
% Physics and Astronomy Classification Scheme (PACS)
\maketitle

\section{introduction}
Topological insulators (TIs) are promising candidates of spintronics materials because of their robust helical surface states and the extremely strong spin-orbit interaction. \cite{1, 2, H. J. Zhang} Initially, binary chalcogenides $Bi_2Te_3$, $Sb_2Te_3$ and $Bi_2Se_3$ \cite{H. J. Zhang} have been identified as three-dimensional TIs by surface sensitive probes such as angle resolved photoemission spectroscopy (ARPES) and scanning tunneling microscopy/spectroscopy (STM/STS). Later, ternary chalcogenide $(Bi_xSb_{1-x})_2Te_3$ \cite{D. S. Kong, J. S. Zhang}, which has similar tetradymite structure to the parent compound $Bi_2Te_3$ and $Sb_2Te_3$, is predicted and confirmed as a tunable topological insulator system to engineer the bulk properties via the Bi/Sb composition ratio with stable topological surface state for the entire composition range by ab initio calculations and ARPES measurements, indicating the robustness of bulk Z$_2$ topology. Combined with magnetism or superconductivity, TIs have attracted great attention due to the rich variety of new physics and applications. Additional spin functionality may occur with the development of ferromagnetism in several transition metal (TM) doped TIs \cite{C. Z. Chang, Y. L. Chen, L. A. Wary, Y. S. Hor, Sb2-xVxTe3, Sb2-xCrxTe3, Y. J. Chien, P. P. J. Haazen, V. K, Y. H. Choi}, which is interesting in itself, because the combination of magnetism with TIs is also good platforms to study fundamental physical phenomena, such as the quantum anomalous Hall effect (QAHE) \cite{N. Nagaosa, X. L. Qi and Y. S. Wu, C. X. Liu, R. Yu, K. Nomura, F. D. M. Haldane}, majorana fermions \cite{Liang Fu and C. L. Kane, Marcel Franz, Jason Alicea, T. Neupert}, image magnetic monopole effect \cite{X. L. Qi and S. C. Zhang}, and topological contributions to the Faraday and Kerr magneto-optical effects \cite{X. L. Qi and T. L. Hughes} etc. topological magnetoelectric effects.

Recently, QAHE has been studied thoroughly and predicted for several TM doped 2D TIs, including HgMnTe quantum wells \cite{C. X. Liu} and thin films of TM (Fe, or Cr) doped $Bi_2Se_3$ \cite{R. Yu}. The quantum anomalous Hall phase derives from the combination of TI and magnetism that breaks the time-reversal symmetry. One common approach to introduce ferromagnetism into a TI is through doping magnetic impurities. Although ferromagnetism has been realized in several TM doped TIs, Cr-doped TIs $(Bi_xSb_{1-x})_2Te_3$ \cite{C. Z. Chang} is the only system that shows the largest anomalous Hall effect (AHE) \cite{N. Nagaosa} until now. Another distinctive character of this system is the carrier-independent ferromagnetism which conflicts with the conventional carrier-mediated mechanism of ferromagnetism in dilute magnetic semiconductor (DMS) \cite{DMS, Dietl RKKY, T. Dietl and H. Ohno, T. Jungwirth}. In current theory, the ferromagnetic exchange interactions between local magnetic moments of dilute magnetic atoms  is mediated by itinerant carriers, so the interaction and Curie temperature T$_c$ are strongly dependent on the distance between magnetic atoms and the density of charge carriers.

In this work, our aim is to further confirm the evolution of ferromagnetism with the change of carrier density and to gain insight into this conflict. The magneto-transport properties of bulk single crystals $Cr_{0.2}Bi_xSb_{1.8-x}Te_3$ are investigated and T$_c$ is found to be proportional to $p^{1/3}$, where p is the hole density. This shows that their bulk magnetic properties can be explained by conventional RKKY mechanism and are very different from their surface ones.

\section{experiments}
$Cr_{0.2}Bi_xSb_{1.8-x}Te_3$ (x = 0, 0.27, 0.36, 0.45, 0.63, 0.90) single crystals were synthesized by melting stoichiometric amounts of high-purity elemental Bi (99.999\%), Te (99.999\%), Sb (99.9999\%) and Cr( 99.95\%) in sealed evacuated quartz glass tube at 800 $^\circ$C for 17 hours, and then slowly cooled down to 550 $^\circ$C by a 24 hours period, and finally naturally cooled to room temperature. The obtained crystals were easily cleaved along the plane with shiny flat surface. The X-ray diffraction (XRD) experiments were carried out with powder samples grinded up from high-quality single crystal pieces at beamline 14B1 of Shanghai Synchrotron Radiation Facility (SSRF), with wavelength 1.2385 {\AA}, divergence angle 2.5 $\ast$ 0.15 mrad$^2$ in horizontal and vertical directions. The MDI Jade 5.0 software was used to analysis the data. Single crystals were used for the observation of magnetic properties using the superconducting quantum interference device (SQUID) magnetometer from Quantum Design. Exfoliated thin flakes of about 85-180 nm thickness on SiO$_2$ were used for magneto-transport experiments by physical property measurement system (PPMS). Both the measurements of temperature dependent longitudinal resistivity $\rho_{xx}$(T) and Hall resistivity $\rho_{xy}$ were carried out in PPMS using the Vander Pauw method with contacts by silver paste and the current was perpendicular to the c axis. Because the estimation of Curie temperature T$_c$ by the appearance of hysteresis or non-zero remnant magnetization is not accurate for disordered systems, the criterion Arrott plots \cite{Arrott} is used here to minimize the effects of magnetic anisotropy and domain rotation. The plots consist in the isotherm curves of H/M versus M$^2$ at different temperatures, where M is the observed magnetization and H the applied magnetic field. A positive or negative slope indicates a second-order or first-order transition, respectively. In Arrott plots, H/M versus M$^2$ should go as  H/M = $a'(T-T_c)+b'M^2+c'M^4$ when H is above coercivity and the intercept changes sign at T$_c$. In our measurements, before each run, the samples were heated above their T$_c$ and cooled to the certain temperature under zero field to ensure perfect demagnetization.

\section{results and discussion}
\subsection{Structure Analysis}
$Sb_2Te_3$ and $Bi_2Te_3$ are narrow band gap semiconductors (E$_g$ = 0.26 eV for $Sb_2Te_3$ and 0.165 eV for $Bi_2Te_3$) with the rhombohedral crystal structure (space group $D^5_{3d}(R\bar{3}m))$ with five atoms in a unit cell and a layered structure stacked along the c axis of the hexagonal lattice. One layer composed of five atoms is known as a quintuple layer (QL) and the interlayer coupling is due to the van der Waals interaction. Crystals of both $Sb_2Te_3$ and $Bi_2Te_3$ grown under stoichiometric conditions are prone to have intrinsic defects, resulting in a typical room temperature natural p-type carriers for $Sb_2Te_3$ and n-type carriers for $Bi_2Te_3$.

\begin{figure}[t]
\includegraphics[width=8.6cm]{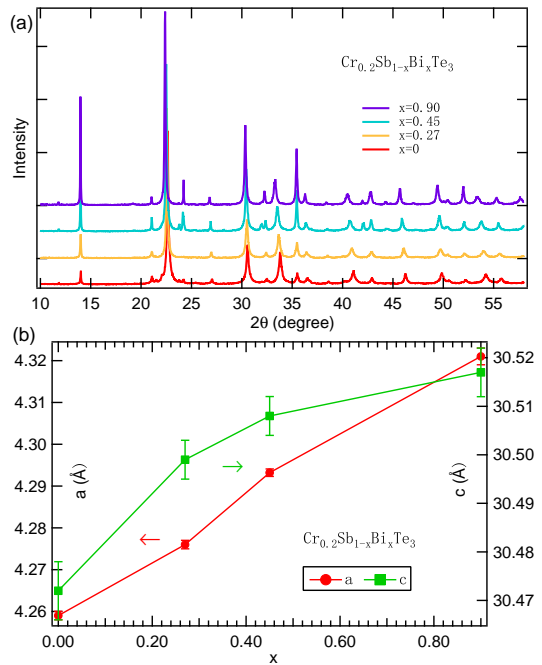}
\caption{(color online) XRD patterns (a) and  the lattice constants a and c (b)  of $Cr_{0.2}Bi_xSb_{1.8-x}Te_3$ as a function of the nominal Bismuth concentration x.}\label{XRD}
\end{figure}

To study the crystal structure, XRD measurements were carried out. According to XRD patterns shown in Fig.\ref{XRD}(a), all XRD peaks of $Cr_{0.2}Sb_{1.8}Te_3$ appear in that of $Cr_{0.2}Bi_xSb_{1.8-x}Te_3$ (x = 0.27, 0.45, 0.9) samples, indicating the unchanged crystal structure of their main phase. To get the lattice constants precisely, two steps are taken to process the data. First determine the locations of peaks by profile fitting, and then refine the lattice constants. The profile fittings are done with linear background in segments, in which Pearson-VII function is used and skewness is set to be zero. The peak locations are used in the cell refinement after the weak peaks and strong overlapping peaks are removed. The obtained lattice parameters a and c are presented in Fig.\ref{XRD}(b). The lattice constants a and c rise with the increase of Bi concentration, showing the smooth substitution of Bi for Sb which can in turn lift Fermi level E$_F$ as reported\cite{D. S. Kong, J. S. Zhang, C. Z. Chang}.

\subsection{Magnetic properties}
\begin{figure}[t]
\includegraphics[width=8.6cm]{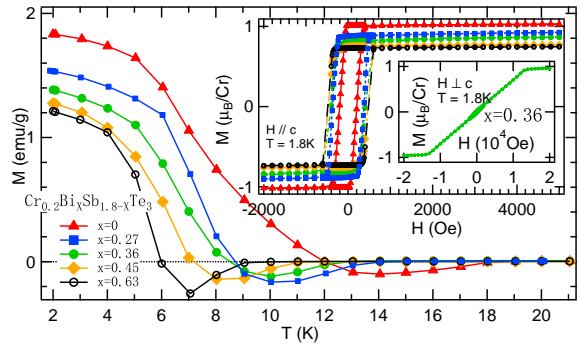}
\caption{(color online) The temperature dependence of remnant magnetization M(T) curves of $Cr_{0.2}Bi_xSb_{1.8-x}Te_3$ after saturated magnetization (H // c, 5 kOe). The insets represent the corresponding M(H) curves measured at 1.8 K for all samples with H // c and for  $Cr_{0.2}Bi_{0.36}Sb_{1.44}Te_3$ only with H $\bot$ c, respectively.}\label{MTH}
\end{figure}

The temperature dependences of remnant magnetization M(T) of $Cr_{0.2}Bi_xSb_{1.8-x}Te_3$ (x = 0, 0.27, 0.36, 0.45, 0.63) after 5 kOe saturated magnetization along c axis are shown in Fig.\ref{MTH} and the Curie temperature $T_c$ can be roughly estimated from these M(T) curves. The inset of Fig.\ref{MTH} represents the corresponding magnetization vs. magnetic field curves at 1.8 K. Compared with the M(H) loop with H $\bot$ c, the sharp, nearly square hysteresis loops with small coercivity with H // c indicate the existence of ferromagnetism with the easy magnetization axis perpendicular to the sample surface. The small coercivity of about 400 Oe (much smaller than that of 12 kOe at 2 K for V-doped $Sb_2Te_3$ \cite{Sb2-xVxTe3}), indicates the magnetic softness of the samples.  The saturated magnetic moment is 0.806 $\mu_B$ per Cr atom for x = 0.36, as seen in the inset of Fig.\ref{MTH}. If the chromium substitutes for bismuth or antimony, it becomes Cr$^{3+}$ with 3d$^3$ electron configuration and its magnetic moment is $\mu_J$ = 0.775 $\mu_B$, so the 0.806 $\mu_B$ saturated magnetic moment shows two facts: almost all Cr atoms are in ferromagnetic order and Cr atoms occupy substituted sites.

The M(H) isotherms and its corresponding Arrott plots of $Cr_{0.2}Bi_{0.36}Sb_{1.44}Te_3$ in the vicinity of its T$_c$ are presented in Fig.\ref{transport}(a) and Fig.\ref{transport}(b), respectively, which indicates that the ferromagnetic transition in  $Cr_{0.2}Bi_{0.36}Sb_{1.44}Te_3$ is of second order as expected, with T$_c$ determined as 11.1 K.

\subsection{Magneto-transport properties}
\begin{figure*}[t]
\includegraphics[width=17.2cm]{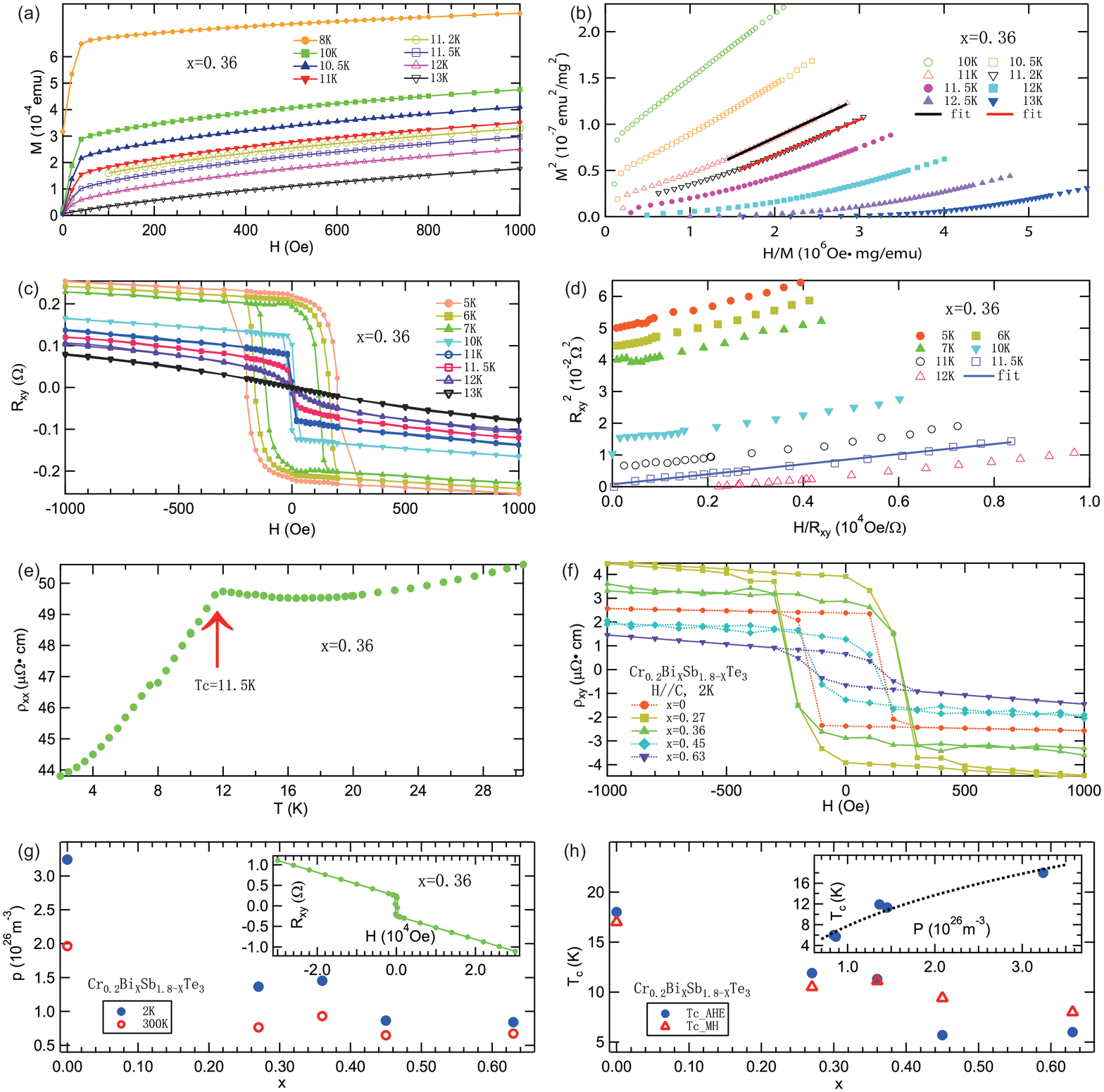}
\caption{(color online) Magneto-transport properties of $Cr_{0.2}Bi_xSb_{1.8-x}Te_3$. The representative results of $Cr_{0.2}Bi_{0.36}Sb_{1.44}Te_3$ sample: the M(H) isotherms (a) and its corresponding Arrott plots (b); $R_{xy}$ isotherms (c) and its corresponding Arrott plots (d); Longitudinal electrical resistivity $\rho_{xx}$(T) (e). (f) The magnetic field dependence of Hall resistivity $\rho_{xy}$ measured at 2 K for $Cr_{0.2}Bi_xSb_{1.8-x}Te_3$ (x = 0, 0.27, 0.36, 0.45, 0.63). (g) The carrier densities at room temperature and 2 K of all samples. Inset: Long range  R$_{xy}(H)$ curve of $Cr_{0.2}Bi_{0.36}Sb_{1.44}Te_3$. (h) Curie temperature T$_c$ with different nominal bismuth concentration x. Inset: Relation betwen T$_c$ and different carrier density at 2K with the theoritical fitting based on RKKY interaction.} \label{transport}
\end{figure*}

Transport measurements were performed of samples with different Bi concentrations and similar results were obtained. Fig.\ref{transport}(a)-(e) and the inset of Fig.\ref{transport}(g) show the representative results of $Cr_{0.2}Bi_{0.36}Sb_{1.44}Te_3$ sample and Fig.\ref{transport}(f)-(h) show the results of $Cr_{0.2}Bi_xSb_{1.8-x}Te_3$ samples with different x.

AHE is a commonly observed phenomenon in ferromagnetic materials caused by spin-orbit interactions, which can be very sensitive to the Berry phase determined by band structure \cite{N. Nagaosa}. It is well known that in ferromagnetic materials the Hall resistivity can be expressed as a sum of two contributions, $\rho_{xy}$ = $\rho_H$ + $\rho_{AH}$ = R$_0$H + R$_s$M = R$_{xy}/d$, where R$_0$ and R$_s$ are the ordinary and the anomalous Hall coefficients, respectively; H is the applied magnetic field; M is the magnetization; d is the thickness of the sample. Anomalous Hall resistivity $\rho_{AH}$ can be deduced by subtracting the ordinary Hall resistivity $\rho_H$ from $\rho_{xy}$. In practice, $\rho_{xy}^{exp}$  contains both longitudinal and Hall contributions, and here the pure Hall contribution $\rho_{xy}$ is extracted from experimental data by the difference of $\rho_{xy}^{exp}$ for positive and negative field directions: $\rho_{xy} = [\rho_{xy}^{exp}(+H) - \rho_{xy}^{exp}(-H)]/2$. This process can remove the $\rho_{xx}$ due to the misalignment of Hall lead, because the background $\rho_{xx}$ is usually symmetric in H.

The $R_{xy}$ of $Cr_{0.2}Bi_{0.36}Sb_{1.44}Te_3$ at different temperatures are shown in  Fig.\ref{transport}(c). The hysteresis loops, namely AHE, further unambiguously confirms its ferromagnetic character. Because Hall resistance R$_{xy}$ reflects magnetization M of ferromagnetic materials, the Arrott plots can also be obtained from R$_{xy}$ measurements.  Fig.\ref{transport} (d) shows the Arrott plots deduced from the data shown in Fig.\ref{transport}(c), in which R$_{xy}^2$ is plotted against H/R$_{xy}$ at each temperature and the extrapolated intercept is proportional to the saturation magnetization M$_s$. The Curie temperature T$_c$ is determined as 11.5K by search the isotherm with zero intercept, almost the same as that determined by SQUID magnetization measurements shown in Fig.\ref{transport}(b). Fig.\ref{transport}(e) shows the longitudinal resistivity $\rho_{xx}$(T) of  $Cr_{0.2}Bi_{0.36}Sb_{1.44}Te_3$. $\rho_{xx}$(T) shows a maximum at about 11.5 K, which is just the Curie temperature, suggesting a paramagnetic to ferromagnetic phase transition here. The similar features have been observed in other DMS, such as $Sb_{2-x}V_xTe_3$ \cite{Sb2-xVxTe3} and $Ga_{1-x}Mn_xAs$ \cite{F. Matsukura} due to the enhanced scattering of carriers by magnetic fluctuation via exchange interaction with localized spins \cite{F. Matsukura}. Besides, $\rho_{xx}$(T) shows a metallic character at the temperature range in paramagnetic phase.
Compared with magnetization measurements that emphasizes local effects, magneto-transport measurements provide an alternative way to study the phase transition, which average over the entire sample. Thus the appearance of peak in longitudinal electrical resistivity $\rho_{xx}$ and AHE below T$_c$ strongly support the presence of broad intrinsic ferromagnetic phase below T$_c$ and the ferromagnetic-paramagnetic phase transition at T$_c$.

The deduced $\rho_{xy}$ of $Cr_{0.2}Bi_xSb_{1.8-x}Te_3$ (x = 0, 0.27, 0.36, 0.45, 0.63) at 2 K are shown in Fig.\ref{transport}(f). The observed $\rho_{AH}$ of above 1 $\mu\Omega\cdot$ cm reflects the large AHE. Given the clear AHE behavior with a T$_c$ from 5 to 18 K in a range easily obtainable and the already  observed quantum Hall effect related to the bulk electronic states in strained bulk HgTe \cite{C. Brune} and $Bi_2Se_3$ \cite{Helin Cao}, $Cr_{0.2}Bi_xSb_{1.8-x}Te_3$ is expected to be a good candidate to observe bulk state related QAHE in the future experiments.

In order to determine the carrier type and density, the ordinary Hall coefficient R$_0$ is determined by the slope of the curve at low temperature and high magnetic field with saturated magnetization. The three-dimensional (3D) holes density p is calculated as p = e/R$_0$, where e is the elementary charge. The inset of Fig.\ref{transport}(g) shows the representative long range  R$_{xy}(H)$ curve of $Cr_{0.2}Bi_{0.36}Sb_{1.44}Te_3$ and the carrier density is observed as 9.3 $\times$ 10$^{25}$ m$^{-3}$ at room temperature and 1.45 $\times$ 10$^{26}$ m$^{-3}$ at 2 K, which are ascribed to the presence of a large number of native anti-site defects. For all samples, negative slopes are observed, which reflect the p-type charge carrier. Fig.\ref{transport}(g) shows the density of holes p of $Cr_{0.2}Bi_xSb_{1.8-x}Te_3$ at room temperature and 2 K, in which p decreases with the increase of Bi concentration x, corresponding to the rise of Fermi level E$_F$ as reported\cite{D. S. Kong, J. S. Zhang, C. Z. Chang}.

\subsection{Ferromagnetism mechanism}
In DMS \cite{DMS}, the density of magnetic impurities is very low compared to the host atoms, thus these magnetic impurities are far apart and which makes the direct coupling between them impossible and the only interaction which can result in ferromagnetism should be carrier-induced via indirect coupling. Ruderman-Kittel-Kasuya-Yoshida (RKKY) interaction is the most famous model dealing with long-range indirect coupling. In RKKY model \cite{Dietl RKKY}, a localized magnetic moment spin-polarizes the conduction electrons and this polarization in turn couples to neighbouring localized moments some distances away. The interaction is long range and has an oscillatory dependence on the distance between the magnetic moments and can be either in ferromagnetic or antiferromagnetic  depending on the separation of the two ions. For a DMS with a typical carrier density, the period of this oscillation becomes very large, and the first zero of the oscillation falls at a distance greater than the cut-off length of the indirect exchange interaction. In this case, the RKKY interaction degenerates into the Zener model developed in 1950. \cite{C. Zener, T. Dietl and H. Ohno}

The carrier-independent ferromagnetism in $Cr_{0.22}(Bi_xSb_{1-x})_{1.78}Te_3$ thin films is observed and explained by the van Vleck mechanism with an intrinsic band origin \cite{C. Z. Chang, R. Yu}, which conflicts with the conventional mechanism of carrier-mediated ferromagnetism in DMS. To study if this unconventional phenomenon is related to the unique topological surface state, the Curie temperatures for different carrier densities are obtained. Figure \ref{transport}(h) summarizes the correlation between nominal Bismuth x and Curie temperature T$_c$. We can clearly see the decrease of  T$_c$ with increase of x. From the dependence of holes concentration p on x at 2K temperature shown in Fig.\ref{transport}(g), the relationship between T$_c$ and p can be established as shown in the inset of Fig.\ref{transport}(h) and can be fitted by T$_c$ $\sim$ p$^{1/3}$ curve with the framework of the RKKY model with mean-field approximation \cite{T. Dietl and H. Ohno, T. Jungwirth}. From this fact, the RKKY interaction is most likely responsible for the appearance of ferromagnetism in $Cr_{0.2}Bi_xSb_{1.8-x}Te_3$ bulk crystal.

\section{conclusion}
Ferromagnetic ordering in $Cr_{0.2}Bi_xSb_{1.8-x}Te_3$ crystal is observed by magnetization and magneto-transport measurements. The Curie temperature $T_c$ is found to depend on hole density and can be explained by a scenario of RKKY interaction. The carrier density dependent character makes the possibility to control the magnetic properties by gate voltage\cite{T. Yokoyama, I. Garate, J. J. Zhu} in the future scientific researches and spintronics applications. The results also show the carrier-independent ferromagnetism in $Cr_{0.22}(Bi_xSb_{1-x})_{1.78}Te_3$ thin films observed by C. Z. Chang et al. \cite{C. Z. Chang} should be related to the topological surface state and the relation between the ferromagnetic interaction and the unique TI surface need to be studied in the future.

\section{acknowledgements}
We acknowledge the technical support and help from stuff members of BL14B1 beamline in SSRF during the XRD measurements. This work is supported by the Natural Science Foundation of China (No: 10979021, 11027401, and 11174054), the Ministry of Science and Technology of China (National Basic Research Program No: 2011CB921800) and Shanghai Municipal Education Commission.

\end{document}